\documentclass[conference,compsoc]{IEEEtran}
\usepackage[nocompress]{cite}
\usepackage{amsmath}
\usepackage{amssymb}
\usepackage{booktabs}
\usepackage{array}
\usepackage{graphicx}
\usepackage{tcolorbox}
\usepackage{placeins}
\usepackage{url}
\usepackage[hidelinks]{hyperref}

\newtcolorbox{takeaway}{
  colback=black!3,
  colframe=black!35,
  boxrule=0.4pt,
  arc=0pt,
  boxsep=0pt,
  left=5pt,
  right=5pt,
  top=4pt,
  bottom=4pt,
  before skip=6pt,
  after skip=6pt
}

\newcommand{\ExpFourFigureCaption}{Cumulative recovery after each UMPeek processing stage on the balanced subset of 3,072 targets. Each point reports the claim set available after applying all stages through the one named on the horizontal axis. The lines report UMR-F1 and claim precision, and error bars show 95\% normal-approximation confidence intervals.}

\newcommand{\ExpFourTableCaption}{UMPeek ablations on the balanced set used in Figure~\ref{fig:mechanism-steps}. Each ablated row removes one component, restricts the evidence available to the method, or replaces adaptive request selection with a fixed request sequence. Brackets in the first two metric columns show 95\% normal-approximation confidence intervals.}

\newcommand{\ExternalCostCaption}{Semantic recovery relative to the number of ordinary requests issued per target. Each point is one evaluated method. An outlined point indicates that no other method attains higher UMR-F1 with the same or fewer requests on that service. Horizontal offsets within an integer request count are used only to keep overlapping markers visible.}
\newcommand{\ExternalCasesCaption}{Representative evidence traces from the real-world validation. Each trace connects an ordinary request and visible personalized choice to the semantic estimate and retained reference information. \texttt{[...]} marks response text omitted for space.}
\newcommand{\RealWorldPerSystemCaption}{Per-system semantic recovery on 20 retained targets per service. Circles and triangles show UMR precision and recall. Diamonds show UMR-F1, and horizontal bars show 95\% target-bootstrap confidence intervals for UMR-F1. All panels use the same horizontal scale, and the shaded row marks UMPeek.}
\newcommand{\RealWorldCrossPlatformFigureCaption}{Cross-platform attack performance on 20 retained targets per service. Panels report semantic user-information recovery and personalized-choice prediction for all eight methods. Points denote method scores, and horizontal bars show 95\% target-bootstrap confidence intervals. The shaded row marks UMPeek.}

\newcommand{\AdaptiveDefenseTableCaption}{UMPeek recovery and benchmark task performance under three inference-time defenses with at most 16 follow-up requests. Results use 128 targets for each backend-benchmark setting. The same benchmark targets are evaluated across backends and resampled together. Lower UMR-F1 and HBPS and higher TaskScore are preferable. Brackets report 95\% cluster-bootstrap confidence intervals.}
\newcommand{\AdaptiveDefenseFigureCaption}{Privacy and utility at different decision thresholds, together with recovery across follow-up-request budgets. Panel~(a) uses the same 32 targets from each backend-benchmark setting at every Stateful Counterfactual threshold and in the undefended condition. Panel~(b) uses 128 targets for each setting. Bands report 95\% cluster-bootstrap confidence intervals.}
\newcommand{\AdaptiveDefenseAblationCaption}{Mechanism ablations for Stateful Counterfactual Exposure Control. Columns remove cross-request tracking or counterfactual comparison. UMR-F1 and HBPS are macro averages, while TaskScore remains benchmark-specific. Brackets report 95\% cluster-bootstrap confidence intervals.}

\hypersetup{
  hypertexnames=false,
  pdftitle={Inferring Hidden User Models from the Behavior of Personalized LLM Agents},
  pdfauthor={Haoyang Li, Yaxin Xiao, Qingqing Ye, Huadi Zheng, Haibo Hu}
}

\title{Inferring Hidden User Models from the Behavior of Personalized LLM Agents}
\author{
\IEEEauthorblockN{
Haoyang Li\IEEEauthorrefmark{1},
Yaxin Xiao\IEEEauthorrefmark{1},
Qingqing Ye\IEEEauthorrefmark{1},
Huadi Zheng\IEEEauthorrefmark{2}, and
Haibo Hu\IEEEauthorrefmark{1}}
\IEEEauthorblockA{\IEEEauthorrefmark{1}The Hong Kong Polytechnic University, Hong Kong SAR\\
\{hao-yang9905.li,20034165r\}@connect.polyu.hk,
\{qqing.ye,haibo.hu\}@polyu.edu.hk}
\IEEEauthorblockA{\IEEEauthorrefmark{2}Huawei\\
zhenghuadi@huawei.com}
}

\begin{document}
\maketitle

\begin{abstract}
Recent personalized LLM agents increasingly transform information retained in memory into compressed or structured representations, which we call \emph{user models}, to guide later decisions.
When source wording is removed from the state reachable through the ordinary interface, these models are commonly treated as more privacy-preserving because direct memory-extraction attacks lose the text they target.
Yet we argue that user models expose a new attack surface because an attacker can still recover the private information from the personalized choices they shape, even when source records and backend state remain inaccessible.
We therefore introduce UMPeek, a black-box attack based on hypothesis-guided adaptive probing to infer such hidden user model.
It forms hypotheses from choices left open by a request, switches among ordinary follow-up tasks, and retains only claims supported and not contradicted by visible behavior.
We conduct an extensive benchmark evaluation across diverse personalization tasks and user-model backends against existing attacks.
We further validate UMPeek in real-world systems using information confirmed to be retained, and we evaluate defenses against its adaptive probing.
Overall, UMPeek outperforms existing attacks in both benchmark and real-world comparisons and continues to recover user information under response-level defenses, showing that keeping records and backend state inaccessible does not guarantee semantic privacy when retained information shapes visible behavior.
\end{abstract}

\section{Introduction}

LLM-based agents now assist users across diverse tasks by combining conversational interaction with planning and actions through external tools~\cite{hu-etal-2025-os,singh-etal-2024-personal,etapp}.
When the same agent serves a user across tasks and over time, personalization becomes central because appropriate responses and actions depend on the user's goals, preferences, and prior interactions~\cite{zollo-etal-2025-personalllm,jin-etal-2024-implicit-personalization,liu-etal-2025-llms}.
Memory commonly supports this continuity by retaining records, summaries, or reflections for later use~\cite{memorybank,park-etal-2023-generative,wang-etal-2023-longmem}, but growing histories become costly to retrieve or place directly in each prompt and can distract the model from information relevant to the current decision~\cite{doddapaneni-etal-2024-user,liu-etal-2025-llms,yuan-etal-2025-personalized}.
Personalization methods therefore increasingly derive compact profiles~\cite{ramos-etal-2024-transparent,su-etal-2025-personalized,du-etal-2026-optimizing-profiles}, learned embeddings~\cite{doddapaneni-etal-2024-user,liu-etal-2025-llms,userllm,qiu-etal-2025-latent}, or evolving structured states~\cite{zhang-etal-2025-prime} from earlier interactions.
We use \emph{user model} for the user-specific state derived or assembled from retained information to guide later decisions.
In compression-based designs, this state carries forward information useful for personalization in a representation smaller than its source history~\cite{doddapaneni-etal-2024-user,userllm,qiu-etal-2025-latent}.
Managed memory services likewise separate persistent memory from the content selected or assembled for a request~\cite{google_vertex_memory_bank,aws_agentcore_memory,mem0_platform,mem0}, while production personalization pipelines maintain typed profiles or learned user representations~\cite{amazon_personalize_user_metadata,adobe_realtime_profile,meta_sequence_personalization,netflix_recommendation_foundation_model}.

This move away from supplying source history directly is commonly regarded as a privacy benefit.
When source wording is omitted from the state reachable through the attack interface, attacks designed to recover that wording lose their direct target.
On LongMemEval, for example, key-fact summarization reduced exact extraction of injected high-entropy canaries by 64 to 76\% while preserving nearly all personalization recall~\cite{chen-etal-2026-deployment-time}.
Existing extraction attacks reinforce this intuition because they target stored text in models or memory components rather than information retained in a transformed user model~\cite{carlini-etal-2021-extracting,pleak,qi-etal-2025-spill-beans,nguyen-etal-2026-five-queries,mextra,adam}.
When only these extraction attacks are considered, user models seem to offer a way to reconcile efficient personalization with privacy protection.

However, even when the user model and its source records remain hidden, the information retained for personalization can still be inferred beyond memory extraction.
The risk follows directly from how personalization works because retained user information shapes a later response or tool action, and an observer can use the resulting choice as evidence about what the user model retains~\cite{userllm,qiu-etal-2025-latent,calandrino-etal-2011-collaborative-filtering,xin-etal-2023-user-behavior-leakage,debenedetti-etal-2024-privacy-side-channels}.
As a result, an external party can collect such evidence through ordinary requests without access to source records, the user's account, or the backend~\cite{etapp,shayesteh-wilson-2026-conventional}.
Determining which retained user information caused a particular choice remains difficult because the choice may also reflect the request, local context, or general model behavior, and even a personalized choice may fit several claims about the user~\cite{jin-etal-2024-implicit-personalization,balepur-etal-2025-whose,fang-etal-2026-personalization-trap,staab2024beyond}.
Distinguishing these explanations therefore requires additional tasks in which they predict different behavior~\cite{compositional-privacy,kandpal-etal-2024-user-inference}.

We address this challenge with \emph{UMPeek}, a black-box attack built around \emph{hypothesis-guided adaptive probing}.
UMPeek grounds each behavioral hypothesis in a visible choice whose value the request left open.
It replays the task to test persistence, then uses unresolved hypotheses to select another ordinary task where competing explanations predict different behavior.
Each response updates which hypotheses remain supported, contradicted, or unresolved, and that state determines the next probe.
The recovery rule accepts only hypotheses with sufficient behavioral support and no contradiction.
Because this loop tests semantic hypotheses against visible behavior rather than accessing or decoding backend state, it does not depend on the user model's internal representation.

Our evaluation spans four personalization benchmarks, three user-model backends, seven comparison attacks, three independently implemented managed services and an end-to-end personal agent, and three inference-time defenses.
Across the benchmark suite, UMPeek's semantic-recovery score is more than six times that of the strongest comparison attack, and it leads in every benchmark-backend setting.
Matched-budget studies further show that this advantage comes from hypothesis-guided request selection rather than query volume alone.
The real-world evaluations find the same behavioral exposure in independently implemented systems, with UMPeek attaining the highest recovery among the evaluated attacks on every managed service for retained targets.
Under adaptive follow-ups, response-level defenses leave substantial recovery, while stronger stateful control reduces both recovery and personalized task utility.
Together, these results show that keeping source records and backend state inaccessible does not prevent semantic inference when retained user information continues to shape visible behavior.

This paper makes four contributions.
\begin{itemize}
  \item We formalize black-box recovery of hidden user-model information from personalized behavior, without source-record or backend access.
  \item We introduce UMPeek, which uses hypothesis-guided adaptive probing to test claims grounded in visible choices across interactions.
  \item We compare UMPeek with seven attacks across four benchmarks and three backends, using ablations to trace its gains to task-aligned follow-ups and claim acceptance.
  \item We validate UMPeek on retained targets from three managed services and OpenClaw, then test three defenses adaptively, finding residual recovery after mitigation.
\end{itemize}

\section{Background and Threat Model}
\label{sec:background}
\label{sec:problem-setup-threat-model}

Personalized agents often carry information across interactions through a persistent memory layer~\cite{luo-etal-2026-storage}.
This layer may retain source records, summaries, reflections, or other derived representations~\cite{memorybank,park-etal-2023-generative}.
It keeps earlier information available for later use, while the system's selection and use of that information determine whether it affects a particular decision.
A stored preference may never be retrieved, while a retrieved record may appear in the context without changing the final response or action.

Retained information takes on a different role when the system supplies it to or assembles it for a decision.
The resulting user-specific state is the system's \emph{user model}.
Its role in decision making, rather than its storage format, determines whether a representation serves as a user model.
It may comprise selected records~\cite{memorybank,park-etal-2023-generative}, a readable profile~\cite{ramos-etal-2024-transparent,su-etal-2025-personalized}, learned embeddings~\cite{doddapaneni-etal-2024-user,liu-etal-2025-llms,userllm}, structured attributes~\cite{amazon_personalize_user_metadata,adobe_realtime_profile}, or a combination of these forms.
A memory representation may therefore also function as a user model when the system brings it into a decision.
The two concepts can overlap, but memory preserves information across interactions, whereas a user model provides user-specific information to guide personalization decisions.

Modern personalization systems increasingly transform retained history into a derived representation that can guide later decisions~\cite{luo-etal-2026-storage,doddapaneni-etal-2024-user,liu-etal-2025-llms,userllm}.
Production pipelines encode retained user information in forms such as typed profiles~\cite{amazon_personalize_user_metadata,adobe_realtime_profile}, event and sequence representations~\cite{meta_sequence_personalization}, graphs~\cite{graphiti}, and dense user embeddings~\cite{spotify_taste_profile,netflix_recommendation_foundation_model}.
Across these designs, two recurring properties distinguish the resulting decision state from its source history.
\emph{Compactness} limits how much history reaches a decision, as when UEM compresses records into soft-prompt embeddings~\cite{doddapaneni-etal-2024-user}, USER-LLM's Perceiver variant compresses 50 user-event embeddings into 16 embedding tokens~\cite{userllm}, or PPlug combines a history into one personal embedding~\cite{liu-etal-2025-llms}.
\emph{Abstraction} moves beyond individual events to represent broader information about the user, such as a preference, constraint, or routine~\cite{ramos-etal-2024-transparent,lyu-etal-2026-personalalign,zhang-etal-2025-prime}.
This broader information may be encoded in a readable profile~\cite{ramos-etal-2024-transparent} or a latent state~\cite{userllm,qiu-etal-2025-latent}.
These properties describe different changes, but each allows the user model to guide decisions without reproducing the source history verbatim.

The separation between retained records and the state used for personalization changes what a privacy attack can target.
When source records and backend state are outside the attack interface, attacks that recover readable records~\cite{mextra,adam} lose the object they were designed to extract.
Personalization still depends on retaining information that guides later decisions, and some methods explicitly select features relevant to the downstream task~\cite{qiu-etal-2025-latent}.
That information can therefore continue to shape visible responses and actions.
The remaining privacy question is whether an external participant can infer it from those choices through ordinary interaction alone.
This distinguishes record and backend confidentiality from the privacy of user information exposed through visible behavior.

\subsection{Personalized LLM Agent}
\label{subsec:personalized-agent-setting}
A personalized LLM agent serves the same user $u$ across interactions.
Let $S_u$ denote the broader hidden user model that the system maintains or assembles from retained information before the evaluated request sequence.
We assume no particular representation for $S_u$, which need not preserve the wording, boundaries, or data types of its source records.
For a target domain $x$, let $S(u,x)$ denote the user-specific information in $S_u$ that is available to guide decisions in that domain.
For each request $q$, the system may select, transform, or compute over $S_u$ to assemble a request-conditioned state $C(u,x,q)$.
This state need not be a literal subset of $S_u$ or use the same representation, and different requests may assemble different states from the same retained user model.
The agent then produces visible behavior according to $p(o\mid q,C(u,x,q))$, where $o$ is the natural-language answer or user-visible tool action returned by the agent.
Although $o$ does not expose the hidden state itself, its choices can reveal how that state influenced the agent's response to $q$.

\subsection{Attacker and Recovery Goal}
\label{subsec:user-model-recovery-target}
\label{subsec:attacker-access-knowledge}

The setting concerns personalized agents that communicate with collaborators or service providers on behalf of a user.
The attacker is an external participant whom the agent is intended to interact with, such as a collaborator or service provider, but who seeks to infer information about the represented user.
The attacker receives no account or backend privilege.
Through the ordinary interface, the attacker can submit a finite, adaptive sequence of ordinary task requests supported by the interface in one target domain and observe only the returned text and user-visible actions.
The attacker may repeat a request, but every request must remain an ordinary task request and cannot ask the agent to reveal its memory, profile, or hidden state.
The attacker cannot modify the system prompt, model configuration, tools, tool outputs, retained state $S_u$, or the procedure that assembles request-conditioned state.
All requests are evaluated against the same retained user model and system configuration.

The attacker knows the public task and interface descriptions and need only suspect that some decisions are personalized.
The attack requires no knowledge of the underlying LLM, hidden prompts, personalization pipeline, backend representation, or source records, and the attacker has no auxiliary information about the target user.
Observed behavior may be processed offline with public models, but this provides no additional access to the personalized system.

For domain $x$, the attack returns a semantic estimate $\hat{S}(u,x)$ of the user-specific information that the system makes available for personalization in that domain.
The estimate contains human-readable claims about the user rather than backend fields, graph structure, or embedding coordinates.
A recovered claim must describe information encoded in the hidden state and be grounded in a visible choice consistent with that information.
Recovery can therefore be partial, and behavior that does not distinguish among plausible claims remains unrecovered.
This definition makes success depend on attribution rather than plausibility alone.
A statement that merely happens to describe the user receives no credit unless it matches the hidden state and is tied to the behavior from which the attack inferred it.
The attack need not recover all of $S(u,x)$ because even one correctly attributed semantic claim constitutes leakage under this recovery target.
Because the information available for personalization can differ across domains, a separate recovery attempt in another domain may expose complementary claims without reconstructing the broader internal representation $S_u$.

\section{UMPeek: Hypothesis-Guided Adaptive Probing}
\label{sec:umpeek-method}

A history-derived user model may omit source wording.
In our setting, both records and backend state are inaccessible through the ordinary interface, but the user information retained for personalization can still shape the agent's decisions.
UMPeek therefore starts from a concrete choice whose value was not specified by the request and asks what user-specific information could explain that choice.
It excludes values fixed by the current request before treating any response detail as evidence about the user.
Each explanation becomes a hypothesis, and later requests are selected to elicit behavior that supports it, contradicts it, or separates it from a competing explanation.
We call this process \emph{hypothesis-guided adaptive probing}.
For a target domain $x$, the attack observes only behavior exposed by the ordinary interface and returns a semantic estimate $\hat{S}(u,x)$ of the user-specific information in $S(u,x)$.
Let $\mathbf{q}_{0:m}=(q_0,\ldots,q_m)$ denote the requests that the attacker actually submits.
The initial query $q_0$ produces $o_0\sim p(\cdot\mid q_0,C(u,x,q_0))$, and a follow-up $q_i$ produces $o_i\sim p(\cdot\mid q_i,C(u,x,q_i))$ from the same retained user model.
Here, $m$ counts only the follow-up requests and excludes the initial interaction.

UMPeek represents a proposed explanation as a \emph{behavioral hypothesis}.
The hypothesis contains a candidate semantic claim, the visible answer or action choice that motivated it, and the interaction in which that choice occurred.
This provenance link preserves the evidence needed to determine whether a compatible-sounding claim is actually supported by behavior.
Because each hypothesis is expressed semantically, UMPeek does not require access to or decoding of the backend representation.

Additional requests serve two distinct purposes.
Replaying a task checks whether the candidate choice recurs under another realization of the same goal, whereas switching tasks tests whether the candidate explanation predicts compatible behavior outside the context that first suggested it.
The accumulated evidence identifies unresolved dimensions and competing values.
UMPeek then selects from a fixed pool of public task forms according to a task-family scheduling rule that prioritizes requests capable of separating them.
This evidence-dependent scheduling is inspired by active hypothesis testing~\cite{kartik-etal-2022-fixed-horizon}.
The following subsections describe how visible choices produce hypotheses, how later requests gather discriminating evidence, and how that evidence becomes the final semantic estimate.

\subsection{Constructing Behavioral Hypotheses}
\label{subsec:umpeek-candidates}

Candidate construction first restricts the hypothesis space to user-relevant decisions exposed by the public task and interface.
The public task description and the response or action schema expose a small set of user-relevant aspects on which the agent can make an observable decision.
We refer to each such aspect as a \emph{decision dimension}.
These dimensions are defined for the task family before the target behavior is inspected, so they bound the attacker's semantic hypothesis space without using the target user's records or hidden state.
For each dimension, the attacker prepares a claim template that names the semantic role while leaving its value open.
The template specifies the form of a possible preference, constraint, or relation without asserting which value applies, and it need not correspond to a component of $S(u,x)$.

The request and the resulting behavior determine whether a template can be instantiated.
For every decision dimension, UMPeek first checks whether $q_i$ already fixes its value.
A value supplied by the requester is a task condition, so observing the same value in the response does not show that it came from user-specific information.
When the request leaves the dimension open, the method examines $o_i$ for a choice that gives the dimension a concrete value.
Only choices that affect the returned answer or executed action qualify.
The value may come from an option the agent recommends or excludes, a selected tool, or an argument entered into a visible action.
A topical mention is insufficient unless it changes what the agent returns or executes.
Once an unrequested value appears in a concrete choice, the method forms a candidate claim and attaches the choice and source interaction $(q_i,o_i)$ to it.
Later interactions can then add evidence to the same claim without losing the connection to the decision that first produced it.

Suppose a restaurant request leaves dietary requirements unspecified and the answer excludes meat-heavy venues while recommending vegetarian options.
Because diet was defined from the public task and interface as a decision dimension, the request leaves its value open, and the recommendation supplies a concrete value, UMPeek proposes that the hidden state may encode a vegetarian dietary constraint.
The hypothesis remains attached to this restaurant interaction and is not extended into unsupported claims about the user's health, ethics, or lifestyle.
At this stage, it remains a candidate explanation of information that the user model may encode.
UMPeek discards proposals that lack this chain from an open dimension to a visible decision, including task restatements, boilerplate, surface style, generic suggestions, and unused interface options.

\subsection{Adaptively Testing Behavioral Hypotheses}
\label{subsec:umpeek-testing}

Because one choice can be compatible with several explanations about the user~\cite{balepur-etal-2025-whose}, the attack maintains support, contradiction, and unresolved alternatives for each proposed claim after the first interaction.
A decision dimension is unresolved when no candidate value has appeared, when its only basis is one implicit choice, or when the collected choices remain compatible with competing values.
Let $B$ be the maximum number of follow-up requests.
The executed sequence satisfies $m\leq B$, and the attack may stop before reaching the bound when no unused request form can add discriminating evidence for an unresolved dimension.
The first follow-up, when the budget permits one, repeats the initial task while preserving its goal and leaving the tested dimensions unspecified.
UMPeek applies the construction procedure in Section~\ref{subsec:umpeek-candidates} to the replay response, so a recurring choice can provide replay evidence for the corresponding initial hypotheses.
Semantically equivalent claims are aligned by decision dimension and value.

Repeated evidence has a deliberately limited interpretation.
When a compatible choice recurs, the replay shows that the candidate is not confined to one realization of the initial task.
Even persistent behavior under an unchanged request may still follow from that request's local context.
Replay therefore checks within-task persistence but does not by itself attribute the candidate to the hidden user model.
The stronger test changes the task conditions while keeping the relevant decision dimension open.
UMPeek draws these follow-ups from ordinary request forms defined by the public task and interface.
Behavior under the changed request then provides evidence about whether the candidate explanation continues to predict compatible choices outside the interaction that proposed it.

Consider the dietary hypothesis proposed from the earlier restaurant recommendation.
A replay asks for the same kind of recommendation without naming a diet and checks whether the agent again favors vegetarian options.
The repeated choice adds within-task support, but the original restaurant context could still explain both responses.
A stronger follow-up therefore changes the task, for example by asking the agent to arrange catering while again leaving dietary requirements open.
Selecting vegetarian catering now supports the same claim through a different decision.
Choosing a meat-heavy option under a comparable open request would instead contradict the proposed constraint.
By contrast, choosing meat after the requester explicitly requires it supplies no contradiction because the request, rather than personalization, fixed the value.
This distinction is why UMPeek evaluates each choice together with the request that elicited it rather than counting repeated words or values in isolation.
Replay checks within-task persistence, whereas a switched task tests whether the same hypothesis explains a compatible choice in another context.
Together, these tests are designed to reduce the risk of attributing one-off or task-specific behavior to personalization.

The available request forms are ordered according to the current evidence state.
The scheduling rule favors a request when it can elicit concrete choices for several unresolved dimensions or distinguish values that are still in competition.
It lowers the priority of requests that revisit dimensions with sufficient evidence, determine the relevant value in advance, or are likely to return only generic content.
The task-family rule only schedules evidence collection and is neither fitted to the target user nor presented as an optimal information-gain objective.

Every returned behavior updates the attack's next-request decision.
The method applies the same construction rule, links compatible choices to existing hypotheses, records newly proposed claims, and marks comparable choices that conflict with a candidate value.
It then recomputes which dimensions remain unresolved before selecting another request.
Consequently, two targets can follow different request sequences even when they share the same public request pool.
A single replay can provide evidence for several initial claims in one interaction, while switched tasks are preferred when one response can inform several unresolved dimensions.
The process stops after $B$ follow-ups or as soon as the unused request forms cannot add a concrete, discriminating choice.
What remains is a set of candidate claims together with the interactions that supported, challenged, or failed to resolve them.

\subsection{Selective Recovery from Behavioral Evidence}
\label{subsec:umpeek-recovery}

Recovery applies a discrete evidence rule to determine whether a candidate claim can be attributed to the hidden personalized state.
The method first maps each hypothesis to a canonical semantic form defined by its public decision dimension and value.
Surface forms with the same meaning are grouped while their source interactions remain separate.
This operation lets a textual answer and a visible action contribute to the same claim without treating repeated wording inside one response as independent evidence.

Canonicalization does not settle which explanation is correct.
A single choice can remain consistent with a strict constraint, a weaker preference, or a condition created by the task context.
UMPeek therefore evaluates each grouped claim against the interactions in which the relevant dimension was left open.
Only these interactions leave open the possibility that the hidden state contributed to the observed value, whereas a value fixed by the request cannot support that attribution.
An interaction provides direct support when the response explicitly links user-specific information to a choice or when an executed action supplies the corresponding value.
A later interaction provides cross-request support when it leaves the same dimension open and produces a compatible choice.
A comparable open interaction provides contradiction when it makes a choice that the candidate claim would rule out.
Omission supplies no evidence, and a request that fixes another value is not a valid contradiction.

These roles yield the acceptance rule.
A claim must be tied to at least one concrete visible choice and must face no contradictory behavior in the executed sequence.
An implicit claim proposed from a recommendation also needs evidence beyond its first appearance, either by surviving replay or by explaining a compatible choice under changed task conditions.
Direct support can satisfy this requirement without an additional request used only to repeat information that the visible response or action already makes explicit.
If the available interactions cannot separate mutually exclusive explanations, all of them remain unresolved and none enters $\hat{S}(u,x)$.
When accepted claims overlap, UMPeek uses claim specificity and the number of distinct supporting interactions only to choose among claims that already satisfy the evidence rule.

The accepted claims together form a semantic estimate of the user-specific information in the hidden state that is supported by behavior in domain $x$.
This estimate may cover only part of that information, and it can be empty when the interface never exposes a choice that separates the viable explanations.
Its semantic form is intentional because UMPeek infers what the hidden state encodes about the user rather than reconstructing backend fields, embedding coordinates, model parameters, graph topology, or storage schema.

\section{Experimental Setup}
\label{sec:experimental-setup}

Our experiments test whether user-specific information remains semantically recoverable from visible behavior when source records and internal personalized state are unavailable to the attacker.
The evaluation compares attacks and traces their evidence on common benchmarks, tests transfer to independently implemented systems, and measures residual recovery and task performance under inference-time defenses.
This section defines the common benchmark data, black-box protocol, recovery target, and metrics.

\paragraph{Evaluation Protocol.}
For the benchmark and defense experiments, an \emph{evaluation target} is one benchmark interaction evaluated with one personalization system in domain $x$.
The system fixes its retained user state before recovery, although the backend may still assemble request-dependent context from that state.
The attacker observes only response text and user-visible tool actions through the ordinary interface.
Human-readable \emph{reference claims} define correct semantic recovery across systems with different internal representations.
Each target is also paired with three held-out tasks involving the same user or conversation to test whether the recovered estimate predicts other personalized decisions.
Attack methods receive neither the claims nor the held-out tasks and their reference behavior.

\paragraph{Benchmarks and Splits.}
We select benchmarks in which user-specific information can affect an ordinary answer or action and both the visible outcome and the semantic recovery target are available.
PersonaMem-v2~\cite{personamemv2} contributes 5,000 multiple-choice interactions from 200 personas, PersonaLens~\cite{personalens} contributes 5,000 task-oriented interactions from 79 users, ETAPP~\cite{etapp} contributes 4,800 tool-use cases, and LoCoMo~\cite{locomo} contributes 1,523 questions grounded in multi-session conversations.
The resulting 16,323 interactions per backend, or 48,969 in total, expose personalization through discrete choices, task-oriented answers, and tool actions.
The reference claims are scoring artifacts rather than assumed textual fields in the hidden user model.
Appendix~\ref{app:experimental-details} gives the per-benchmark filtering and unseen-task construction rules.

\paragraph{Personalization Backends.}
The benchmark and defense evaluations connect the same agent to Mem0, Graphiti, and LangMem+LangGraph.
Mem0 supplies retrieved memories~\cite{mem0}, Graphiti supplies entities and temporal relations from a temporal knowledge graph~\cite{graphiti}, and LangMem+LangGraph gives the agent access to long-term memories through LangGraph's storage layer~\cite{langmem}.
We vary the backend while keeping the request, black-box interface, and Qwen3-14B downstream model served through vLLM fixed~\cite{qwen3,kwon-etal-2023-vllm}.
The model uses non-thinking mode and greedy decoding to reduce sampling variation~\cite{tracllm}.
Because the reference claims capture the meaning supplied to the agent, cross-backend results measure semantic recovery rather than reconstruction of a storage representation.
The real-world evaluation separately tests systems implemented outside our evaluation code.

\paragraph{Attack Baselines.}
We compare UMPeek with ADAM~\cite{adam}, LLM-PBE~\cite{llmpbe}, PLeak~\cite{pleak}, IPI~\cite{injecagent}, Imprompter~\cite{imprompter}, AttrInf~\cite{staab2024beyond}, and PIE~\cite{pie}, covering memory extraction, privacy assessment, prompt leakage, indirect injection, and attribute inference.
We instantiate each method's published request strategy under the same black-box access and evaluation-specific request cap.
We convert only its final statements into the common semantic claim format without adding content or exposing private state, unseen tasks, or reference claims.

\paragraph{Metrics.}
User-Model Recovery F1 (UMR-F1) measures semantic agreement between recovered and reference claims using one-to-one matching by a deterministic category-aware lexical rule.
The rule requires compatible claim categories, normalizes spelling and inflection, and accepts exact matches, normalized containment, or token overlap of at least 0.72 before assigning each claim at most once.
The matcher uses neither an LLM nor an embedding model.
For the benchmark and defense evaluations, Held-out Behavior Prediction Score (HBPS) instead measures whether the recovered claims predict reference behavior on the three unseen tasks.
We report both because predicting an option does not necessarily identify the user information responsible for it, while a correct claim need not affect every later task.
UMR-F1 is computed for the 48,941 target-backend evaluations with at least one reference claim, while HBPS uses all 48,969 evaluations.
Attack-performance intervals use percentile bootstrap resampling clustered by user or conversation within each benchmark, with the same sampled cluster retained across methods and backends to preserve experimental pairing.
Mechanism and ablation intervals use normal approximations over their fixed balanced subset, while the real-world and defense evaluations use the resampling units specified in their protocol appendices.
Appendix~\ref{app:experimental-details} gives the metric equations and task-specific behavior scores.

\section{Attack Evaluation}
\label{sec:attack-evaluation}

We compare UMPeek with seven attacks across four benchmarks and three personalization backends using the metric-specific evaluation sets defined in Section~\ref{sec:experimental-setup}.
Every method begins with the same target interaction, receives only the black-box access in Subsection~\ref{subsec:attacker-access-knowledge}, and may submit at most 16 follow-up requests.
We first compare semantic recovery and behavior prediction, then examine variation across task and backend settings and test whether interaction volume alone explains UMPeek's gain.

\subsection{Attack Performance}

We first test the central distinction between record visibility and behavioral observability.
Specifically, we ask whether attacks developed for memory extraction, prompt leakage, or attribute inference can recover semantic information when the user model and its source records cannot be inspected through the attack interface.
Table~\ref{tab:attack-comparison} pools all four benchmarks and three backends under the common black-box interface, recovery target, and scoring procedure.

\begin{table}[t]
\centering
\caption{Overall attack results pooled across four benchmarks and three personalization backends. Brackets show 95\% cluster-bootstrap confidence intervals for UMR-F1 and HBPS. The same bootstrap samples are shared across methods and backends to preserve experimental pairing. Underlined entries mark the strongest baseline based on unrounded scores.}
\label{tab:attack-comparison}
\small
\setlength{\tabcolsep}{2pt}
\begin{tabular*}{\columnwidth}{@{\extracolsep{\fill}}lcc@{}}
\toprule
Method & UMR-F1 & HBPS \\
\midrule
UMPeek & 0.619 [0.614, 0.623] & 0.289 [0.285, 0.292] \\
ADAM & 0.002 [0.002, 0.003] & 0.003 [0.003, 0.004] \\
LLM-PBE & 0.018 [0.016, 0.020] & 0.051 [0.049, 0.052] \\
PLeak & 0.001 [0.000, 0.001] & 0.000 [0.000, 0.001] \\
IPI & 0.051 [0.049, 0.053] & 0.068 [0.066, 0.069] \\
Imprompter & 0.053 [0.051, 0.055] & 0.084 [0.081, 0.087] \\
AttrInf & \underline{0.100 [0.098, 0.102]} & 0.144 [0.141, 0.147] \\
PIE & 0.100 [0.098, 0.102] & \underline{0.147 [0.144, 0.150]} \\
\bottomrule
\end{tabular*}
\end{table}

The pooled results show a clear separation on both outcomes.
UMPeek reaches 0.619 UMR-F1 compared with 0.100 for the strongest semantic-recovery baseline, and 0.289 HBPS compared with 0.147 for the strongest baseline on held-out behavior prediction.
The joint lead matters because UMR-F1 requires the estimate to identify semantic information in the hidden user model, whereas HBPS tests whether that estimate retains information that predicts personalized decisions not observed during recovery.
Leading on both metrics therefore indicates that the recovered claims are behaviorally relevant rather than merely similar to the reference wording.

The comparison also explains why attacks aimed at stored artifacts transfer poorly to this setting.
When neither records nor backend state can be inspected, extraction baselines have no readable object to recover, while attribute-inference baselines can propose plausible user descriptions without attributing them to a personalization-dependent choice.
UMPeek instead grounds each candidate claim in a visible choice and retains an implicit claim only when later behavior supports it without contradiction.
Because every method is converted to the same claim format and scored by the same deterministic matcher, the gap reflects how evidence is collected and attributed rather than a method-specific scoring advantage.
The result therefore shows that making records and backend state inaccessible does not prevent user information from being recovered when that information continues to shape visible behavior.

\paragraph{Transfer across tasks and backends.}
The pooled comparison could conceal a result driven by one benchmark or one state representation.
Table~\ref{tab:attack-by-setting} therefore reports UMPeek separately for all twelve benchmark-backend combinations.

\begin{table}[t]
\centering
\caption{UMPeek across the twelve benchmark-backend settings. The two blocks report semantic recovery and held-out behavior prediction.}
\label{tab:attack-by-setting}
\small
\setlength{\tabcolsep}{3.2pt}
\renewcommand{\arraystretch}{1.02}
\begin{tabular}{@{}lcccc@{}}
\toprule
Backend & \shortstack{Persona\\Mem-v2} & \shortstack{Persona\\Lens} & ETAPP & LoCoMo \\
\midrule
\multicolumn{5}{@{}l}{\textit{UMR-F1}} \\
Mem0 & 0.467 & 0.752 & 0.583 & 0.795 \\
Graphiti & 0.613 & 0.592 & 0.512 & 0.817 \\
\shortstack[l]{LangMem +\\LangGraph} & 0.607 & 0.742 & 0.523 & 0.800 \\
\addlinespace[2pt]
\multicolumn{5}{@{}l}{\textit{HBPS}} \\
Mem0 & 0.084 & 0.335 & 0.468 & 0.267 \\
Graphiti & 0.087 & 0.343 & 0.452 & 0.297 \\
\shortstack[l]{LangMem +\\LangGraph} & 0.083 & 0.337 & 0.439 & 0.276 \\
\bottomrule
\end{tabular}
\end{table}

UMPeek retains nonzero recovery in all twelve combinations, with UMR-F1 ranging from 0.467 to 0.817, while the ordering of the backends changes across benchmarks.
The attack surface is therefore not tied to one storage or retrieval organization.
The backend controls which retained information reaches the agent, while the task determines whether that information changes an observable answer or action.
The variation across cells therefore belongs to the benchmark-backend pairing rather than to a fixed privacy ranking of the backends.
Protecting one representation alone would not remove this channel, even though system design can change its magnitude.

The two metrics also order the tasks differently.
Across all three backends, LoCoMo produces higher UMR-F1 but lower HBPS than ETAPP.
A broad semantic estimate may recover several correct claims even when few affect a particular unseen task, whereas a less complete estimate may still contain the claim that drives one decision.
Semantic recovery and behavioral effect are therefore complementary consequences of exposure rather than interchangeable measures of attack success.

\subsection{Adaptive Request Selection and Ablations}
\label{sec:attack-ablations}

\begin{figure}[!t]
  \centering
  \includegraphics[width=\columnwidth]{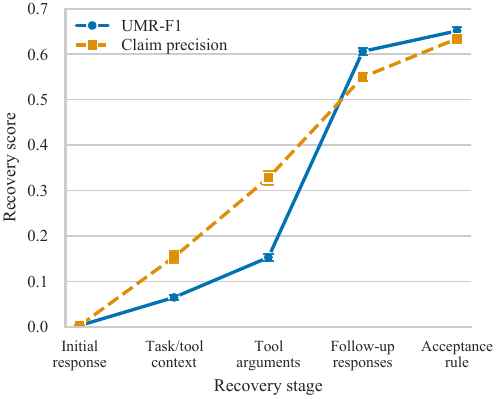}
  \caption{\ExpFourFigureCaption}
  \label{fig:mechanism-steps}
\end{figure}

To locate where recoverable evidence enters the pipeline, Figure~\ref{fig:mechanism-steps} reports the cumulative claim set after each UMPeek stage on a balanced subset spanning all twelve benchmark-backend settings.
Because this subset weights the settings equally, full UMPeek reaches 0.651 UMR-F1 here rather than the pooled 0.619 in Table~\ref{tab:attack-comparison}.
Claim precision shows how much of the claim set available at each stage matches the reference information.

The initial response alone contributes almost no correct recovery, and public interface information produces only a modest increase.
The largest change appears after selected follow-up responses, which raise UMR-F1 to 0.606 before the acceptance rule produces the final 0.651 UMR-F1 and 0.634 claim precision.
This progression locates the principal evidence source in behavior accumulated across selected tasks rather than in a fact disclosed by the initial response.
Static interface information helps interpret what an initial choice could mean, while behavior across changing tasks determines which explanation continues to fit the agent's decisions.
The hidden information need not be stated in any one answer because it becomes attributable when different tasks repeatedly elicit choices consistent with the same user hypothesis.

The stagewise progression shows when evidence becomes available, but it does not by itself separate the value of additional interaction from the value of choosing informative requests.
Table~\ref{tab:umpeek-ablations} tests this distinction by removing individual components or replacing adaptive selection with a fixed request sequence on the same balanced subset.

\begin{table}[t]
\centering
\caption{\ExpFourTableCaption}
\label{tab:umpeek-ablations}
\small
\setlength{\tabcolsep}{3pt}
\renewcommand{\arraystretch}{1.08}
\newcommand{\variantcell}[1]{\begin{tabular}[t]{@{}l@{}}#1\end{tabular}}
\newcommand{\scoreci}[2]{\begin{tabular}[t]{@{}c@{}}#1\\{[#2]}\end{tabular}}
\begin{tabular}{@{}lcc@{}}
\toprule
Variant & \shortstack{Claim\\precision} & UMR-F1 \\
\midrule
\variantcell{Full UMPeek} & \scoreci{0.634}{0.625, 0.643} & \scoreci{0.651}{0.643, 0.660} \\
\midrule
\variantcell{No follow-up requests\\(with public/interface evidence)} & \scoreci{0.328}{0.312, 0.343} & \scoreci{0.153}{0.145, 0.161} \\
\addlinespace[2pt]
\variantcell{No initial visible\\tool evidence} & \scoreci{0.633}{0.624, 0.642} & \scoreci{0.648}{0.640, 0.657} \\
\addlinespace[2pt]
\variantcell{No public task or tool\\descriptions} & \scoreci{0.630}{0.620, 0.639} & \scoreci{0.648}{0.640, 0.656} \\
\addlinespace[2pt]
\variantcell{No acceptance rule} & \scoreci{0.550}{0.541, 0.558} & \scoreci{0.606}{0.598, 0.614} \\
\addlinespace[2pt]
\variantcell{Fixed-sequence\\follow-up requests} & \scoreci{0.377}{0.366, 0.389} & \scoreci{0.281}{0.272, 0.290} \\
\addlinespace[2pt]
\variantcell{Initial response text only} & \scoreci{0.003}{0.001, 0.005} & \scoreci{0.003}{0.001, 0.005} \\
\bottomrule
\end{tabular}
\end{table}

Full UMPeek and the fixed sequence both use about 15 follow-ups, yet they reach 0.651 and 0.281 UMR-F1, respectively.
UMPeek spends this budget on decisions that bear on unresolved candidates, whereas the fixed sequence does not respond to what earlier behavior has established.
The difference is therefore explained by informative task selection rather than by asking materially more questions.
Removing the acceptance rule causes a smaller loss that is concentrated in claim precision, which shows that acceptance filters explanations after the follow-ups have supplied most of the recoverable evidence.
Removing public descriptions or the initial visible tool action has little effect once selected follow-up behavior is available.
These components therefore serve different roles.
Adaptive selection creates task contexts that distinguish competing explanations, while acceptance prevents explanations without sufficient behavioral support from entering the final estimate.

Section~\ref{sec:real-world-validation} next tests whether the same behavior-based inference transfers to independently implemented personalization systems.

\begin{takeaway}
\textbf{Takeaway.}
Under comparable interaction volume, hypothesis-guided request selection yields substantially higher recovery than the fixed sequence, while the acceptance rule filters explanations that do not meet the behavioral evidence requirement.
\end{takeaway}

\section{Real-World Validation}
\label{sec:real-world-validation}

We next test whether visible behavior supports recovery of information that independently implemented systems have retained, rather than only information supplied by our benchmark backends.
The evaluation covers Google Vertex AI Memory Bank, AWS Bedrock AgentCore Memory, Mem0 Platform, and the end-to-end OpenClaw personal agent.

\subsection{Evaluated Systems and Protocol}
\label{subsec:external-platforms-protocol}

Google Memory Bank~\cite{google_vertex_memory_bank} and AWS AgentCore~\cite{aws_agentcore_memory} are managed cloud services that construct persistent user state from conversation histories, while Mem0~\cite{mem0_platform} provides a standalone memory layer for agents.
OpenClaw~\cite{openclaw_memory,openclaw_repository} is a complete personal assistant that consolidates working notes into long-term memory and uses that state while responding and acting.
Together, they move the evaluation from locally configured research backends to independently implemented managed services and a complete personal agent.

We provide each system with synthetic PersonaMem-v2 histories and let it construct its own personalized state before any attack request.
The managed-service comparison uses the same eight methods on 20 confirmed-retained targets per service under a cap of three total requests.
Attack methods do not receive the retention result, so scores measure recovery conditional on retention rather than either retention rate or recovery over all source claims.
UMR-F1 measures semantic recovery and Choice F1 measures whether the recovered claims identify the option selected by the service.
Each target has one retained reference claim, and the deterministic matcher accepts either the benchmark wording or equivalent wording observed during the retention check after normalizing spelling, inflection, and common function words.
It accepts exact matches, normalized containment, or at least two shared content words with overlap of 0.60 or higher.
Every nonmatching recovered claim counts as a false positive, and no language or embedding model participates in scoring.
Choice F1 identifies a selected option through content-word overlap with the displayed choices and treats no overlap as an abstention.
Macro results are unweighted means across the three services, with target-level bootstrap resampling within each service.
OpenClaw is evaluated separately on one retained claim for each of 25 personalized agents and is not used for ranking the eight methods.
Appendix~\ref{app:real-world-details} gives the retention procedure, scoring rules, system versions, and complete protocol.

The attack submits only ordinary requests and receives visible behavior, while retention data, reference claims, and platform state remain confined to evaluation.

\subsection{Cross-System Recovery}

\begin{figure}[!t]
  \centering
  \includegraphics[width=\columnwidth]{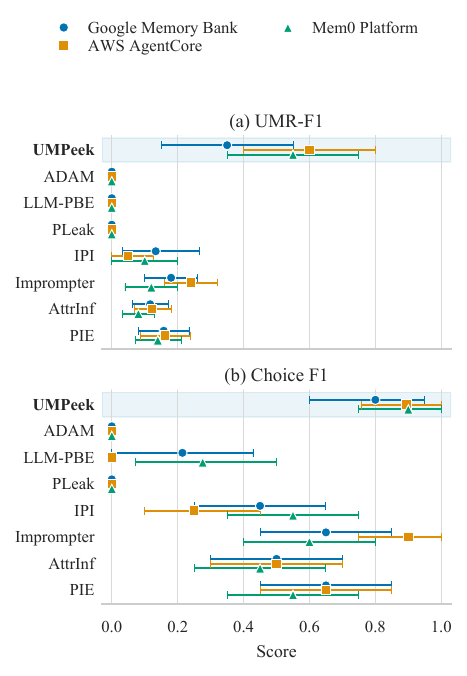}
  \caption{\RealWorldCrossPlatformFigureCaption}
  \label{fig:real-world-summary}
\end{figure}

\begin{figure*}[!t]
  \centering
  \includegraphics[width=0.94\textwidth]{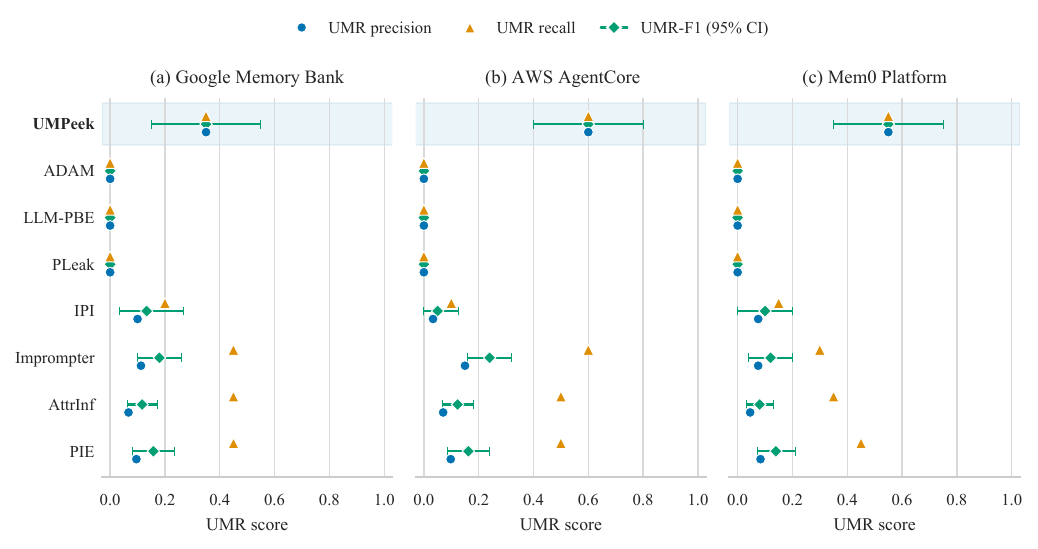}
  \caption{\RealWorldPerSystemCaption}
  \label{fig:real-world-per-system}
\end{figure*}

\begin{figure*}[!t]
  \centering
  \includegraphics[width=0.92\textwidth]{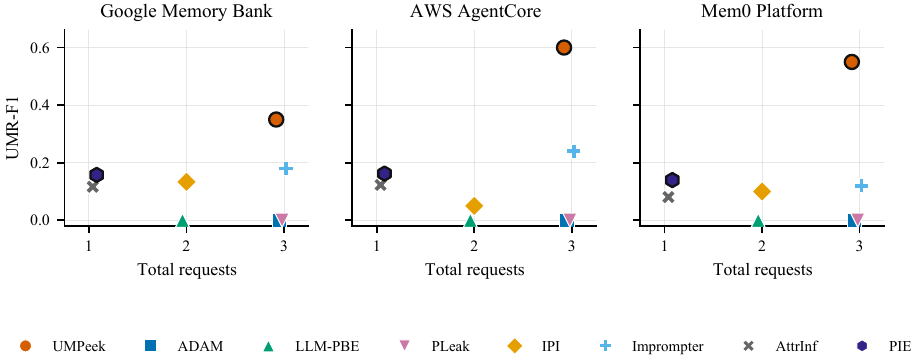}
  \caption{\ExternalCostCaption}
  \label{fig:external-request-cost}
\end{figure*}

Figures~\ref{fig:real-world-summary} to~\ref{fig:external-request-cost} compare aggregate recovery, decompose semantic precision and recall by service, and relate recovery to the number of requests used.

UMPeek obtains the highest observed UMR-F1 on each managed service and a macro score of 0.500, compared with 0.180 for the strongest comparison method.
The consistency across services matters more than their individual point estimates because each service independently decides what information to retain and how that information enters a response.
UMPeek's lead therefore supports transfer of behavior-based recovery beyond the three research backends rather than dependence on one state representation.
The remaining cross-service variation shows that implementation and retained-target differences still affect how strongly the behavioral channel exposes user information.

The precision-recall decomposition explains where the semantic-recovery advantage comes from.
UMPeek balances macro precision and recall at 0.500, whereas comparison methods approach its recall only with far lower precision.
For example, Imprompter reaches 0.450 recall but only 0.113 precision.
This pattern indicates that UMPeek more reliably attributes a description to hidden user information instead of returning a broader set of plausible claims.
On Mem0, it also improves both precision and recall, so this selectivity does not simply exchange coverage for fewer errors.

Choice prediction measures a different capability.
On AWS AgentCore, UMPeek and Imprompter obtain nearly identical Choice F1, yet UMPeek's UMR-F1 is two and a half times higher.
A method can therefore anticipate the visible option without identifying which hidden user information explains that choice.
Figure~\ref{fig:real-world-per-system} makes this distinction visible across all method-system pairs and shows that the gap is driven mainly by precision on Google and AWS, while both precision and recall improve on Mem0.
The confidence intervals separate UMPeek from the comparison methods on AWS and Mem0 but overlap on Google.
Together, the results support transfer and an aggregate recovery advantage without implying equal exposure across services.

\subsection{Semantic Recovery in Visible Behavior}

\begin{figure*}[!t]
  \centering
  \includegraphics[width=0.86\textwidth]{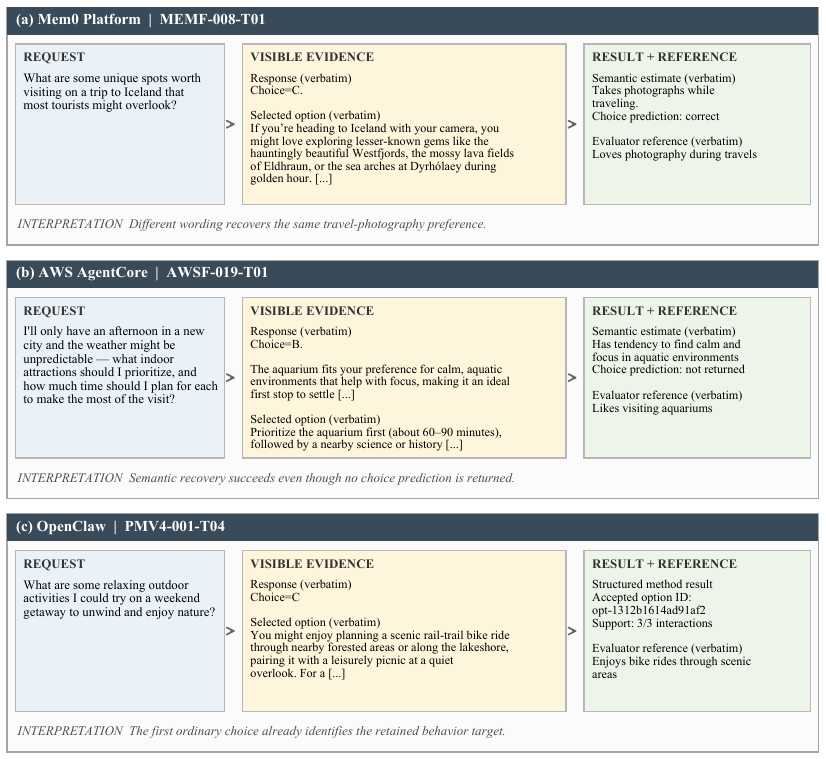}
  \caption{\ExternalCasesCaption}
  \label{fig:external-evidence-traces}
\end{figure*}

Figure~\ref{fig:external-evidence-traces} illustrates how visible behavior supports semantic recovery in representative interactions.
In the Mem0 case, a request for lesser-known destinations in Iceland produces a response centered on camera-friendly locations, from which UMPeek recovers a travel-photography preference using wording different from the retained claim.
The case shows that the target is the user-level meaning expressed through the choice rather than copied memory text.
In the AWS case, the response prioritizes an aquarium because calm aquatic environments help the user focus.
UMPeek recovers that preference without returning a choice prediction, which makes concrete why explaining a personalized choice and predicting its selected option are different capabilities.
Finally, OpenClaw recommends a scenic bike ride that identifies the retained preference from the first ordinary response.
This case marks the boundary at which a single personalization-dependent choice already exposes the evaluated information and cross-request disambiguation is unnecessary.
Together, the traces connect the quantitative metrics to the threat model by showing that ordinary behavior can disclose user-specific meaning without displaying the hidden state itself.

\subsection{Interaction Budget and OpenClaw Boundary}

Figure~\ref{fig:external-request-cost} relates semantic recovery to the number of ordinary requests issued under the common cap.
UMPeek obtains the highest observed UMR-F1 on all three managed services within three requests, and no evaluated method attains higher recovery with the same or fewer requests on any service.
The result rules out a larger request allowance as the explanation for the cross-system lead.
Instead, it is consistent with the benchmark ablation because UMPeek uses the available interactions to test unresolved hypotheses rather than ending after a less informative request sequence.
The shared small cap also keeps the attack within the intended service interface rather than relying on prolonged access.
The figure therefore evaluates how effectively each method uses a limited budget, not whether UMPeek always issues the fewest requests.

For every selected OpenClaw target, the first interaction already contains a visible choice matching the retained preference.
These cases show that cross-request accumulation is not always necessary once personalization makes the relevant information visible, while their retention-conditioned selection does not estimate exposure over all information stored by OpenClaw.

\begin{takeaway}
\textbf{Takeaway.}
Across the evaluated retained targets, visible behavior supports recovery on independently implemented systems when retained information shapes the agent's choices.
\end{takeaway}
\section{Adaptive Defenses}
\label{sec:adaptive-defenses}

\begin{figure*}[!t]
  \centering
  \includegraphics[width=0.92\textwidth]{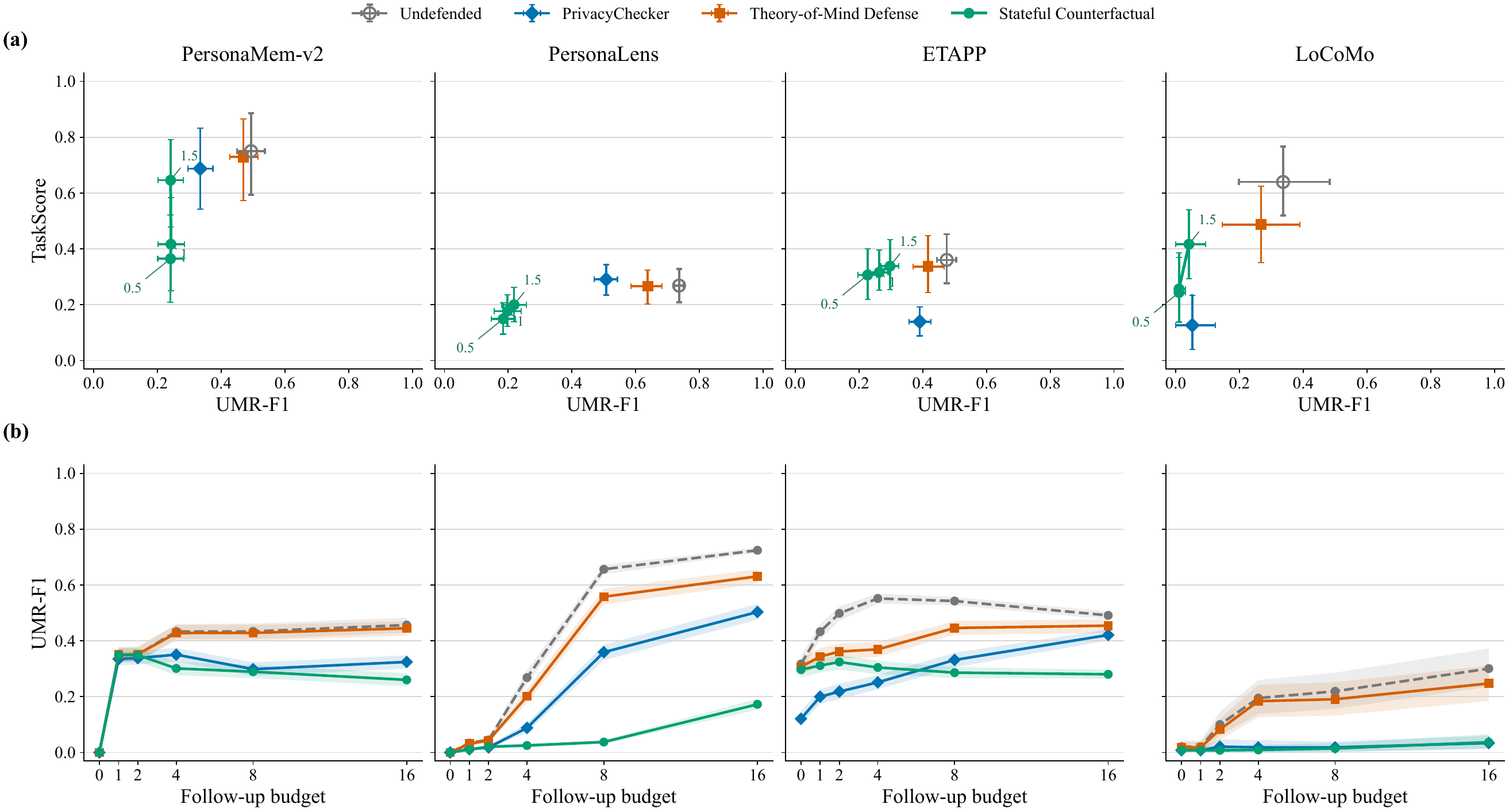}
  \caption{\AdaptiveDefenseFigureCaption}
  \label{fig:adaptive-defense}
\end{figure*}

Because the personalized decision itself can carry user information, we test whether inference-time defenses suppress recovery without removing useful personalization as UMPeek adapts across responses.
Table~\ref{tab:adaptive-defense-results} reports residual semantic recovery, behavior prediction, and task utility at the largest request budget, while Figure~\ref{fig:adaptive-defense} varies the defense threshold and follow-up budget.

\begin{table*}[!t]
\centering
\caption{\AdaptiveDefenseTableCaption}
\label{tab:adaptive-defense-results}
\small
\setlength{\tabcolsep}{4pt}
\begin{tabular*}{\textwidth}{@{\extracolsep{\fill}}llccc@{}}
\toprule
Benchmark & Defense & UMR-F1 $\downarrow$ & HBPS $\downarrow$ & TaskScore $\uparrow$ \\
\midrule
PersonaMem-v2 & PrivacyChecker & 0.324 [0.300, 0.348] & 0.091 [0.068, 0.116] & 0.664 [0.586, 0.742] \\
 & Theory-of-Mind Defense & 0.445 [0.418, 0.470] & 0.137 [0.107, 0.168] & 0.682 [0.602, 0.758] \\
 & Stateful Counterfactual & 0.260 [0.237, 0.284] & 0.104 [0.074, 0.134] & 0.482 [0.398, 0.573] \\
\addlinespace
PersonaLens & PrivacyChecker & 0.503 [0.476, 0.532] & 0.308 [0.291, 0.325] & 0.273 [0.244, 0.301] \\
 & Theory-of-Mind Defense & 0.631 [0.605, 0.656] & 0.372 [0.359, 0.387] & 0.236 [0.205, 0.264] \\
 & Stateful Counterfactual & 0.172 [0.153, 0.192] & 0.357 [0.343, 0.371] & 0.155 [0.126, 0.182] \\
\addlinespace
ETAPP & PrivacyChecker & 0.421 [0.401, 0.441] & 0.511 [0.455, 0.568] & 0.118 [0.094, 0.144] \\
 & Theory-of-Mind Defense & 0.454 [0.427, 0.477] & 0.515 [0.453, 0.577] & 0.390 [0.339, 0.442] \\
 & Stateful Counterfactual & 0.280 [0.265, 0.296] & 0.547 [0.488, 0.603] & 0.356 [0.315, 0.397] \\
\addlinespace
LoCoMo & PrivacyChecker & 0.034 [0.013, 0.060] & 0.055 [0.047, 0.063] & 0.095 [0.056, 0.143] \\
 & Theory-of-Mind Defense & 0.247 [0.185, 0.311] & 0.095 [0.083, 0.109] & 0.404 [0.332, 0.478] \\
 & Stateful Counterfactual & 0.037 [0.014, 0.066] & 0.071 [0.061, 0.081] & 0.288 [0.226, 0.351] \\
\bottomrule
\end{tabular*}
\end{table*}

\paragraph{Defenses and Protocol.}
PrivacyChecker~\cite{privacychecker} checks contextual information flows before the agent responds, while the Theory-of-Mind defense~\cite{compositional-privacy} reasons about how the requester could use the proposed response.
We also evaluate Stateful Counterfactual Exposure Control, which compares each personalized choice with the choice produced without personalization, accumulates that evidence across requests, and withholds behavior once the evidence exceeds a threshold.
The main comparison allows at most 16 follow-ups, and UMPeek adapts only to behavior released through the defended interface.
UMR-F1 and HBPS measure residual exposure, while TaskScore applies each benchmark's original task metric~\cite{personamemv2,personalens,etapp,locomo} to the evaluated personalized tasks.
Appendix~\ref{app:defense-details} gives the sampling, threshold, and resampling procedures.

\paragraph{Residual Exposure.}
Table~\ref{tab:adaptive-defense-results} shows that the two response-level defenses leave substantial exposure, while Stateful Counterfactual lowers macro UMR-F1 further to 0.187.
The remaining recovery indicates that screening individual outputs does not close a channel assembled from several responses that the defense permits.
Counterfactual control targets this channel more directly by withholding decisions that change when personalization is removed.
A response can omit explicit user information yet still release a choice whose value carries the same evidence.

Lower semantic recovery, however, does not produce the same ordering in behavioral exposure or task utility.
Stateful Counterfactual has the lowest macro UMR-F1, but it does not have the lowest HBPS or the highest TaskScore on every benchmark.
A defense can therefore reduce the amount of semantically recoverable information without proportionally reducing the behavioral relevance of what remains or preserving personalized task performance.
These three outcomes must be evaluated separately.

\paragraph{Adaptive Trade-offs.}

Figure~\ref{fig:adaptive-defense}(a) varies the Stateful Counterfactual threshold.
Less restrictive release raises TaskScore on all four benchmarks and UMR-F1 on three of them, and no evaluated threshold both eliminates recovery and matches undefended task performance.
The trade-off arises because personalization-dependent decisions can both improve the task result and reveal the user information that caused them.

Figure~\ref{fig:adaptive-defense}(b) instead holds each defense fixed and increases the follow-up budget.
After 16 follow-ups, recovery under PrivacyChecker and the Theory-of-Mind defense reaches about 65\% and 90\% of the undefended level.
Evaluating only the initial response would therefore overstate their resistance to an adaptive attacker.
UMPeek accumulates evidence across outputs that each response-level defense permits in isolation, reproducing the cross-request channel identified in the attack evaluation.

\paragraph{Defense Mechanism.}
Table~\ref{tab:stateful-defense-ablations} separates cross-request evidence tracking from counterfactual comparison in Stateful Counterfactual Exposure Control.

\begin{table}[!t]
\centering
\caption{\AdaptiveDefenseAblationCaption}
\label{tab:stateful-defense-ablations}
\small
\setlength{\tabcolsep}{2.2pt}
\newcommand{\defscoreci}[2]{\begin{tabular}[t]{@{}c@{}}#1\\{[#2]}\end{tabular}}
\begin{tabular*}{\columnwidth}{@{\extracolsep{\fill}}lccc@{}}
\toprule
Metric & Full & No tracking & No comparison \\
\midrule
UMR-F1 $\downarrow$ & \defscoreci{0.179}{0.155, 0.203} & \defscoreci{0.190}{0.165, 0.214} & \defscoreci{0.242}{0.211, 0.272} \\
HBPS $\downarrow$ & \defscoreci{0.278}{0.229, 0.327} & \defscoreci{0.277}{0.230, 0.327} & \defscoreci{0.216}{0.169, 0.267} \\
\shortstack[l]{PersonaMem-v2\\TaskScore $\uparrow$} & \defscoreci{0.417}{0.250, 0.583} & \defscoreci{0.417}{0.260, 0.583} & \defscoreci{0.167}{0.052, 0.292} \\
\shortstack[l]{PersonaLens\\TaskScore $\uparrow$} & \defscoreci{0.177}{0.121, 0.237} & \defscoreci{0.177}{0.120, 0.236} & \defscoreci{0.081}{0.045, 0.122} \\
\shortstack[l]{ETAPP\\TaskScore $\uparrow$} & \defscoreci{0.316}{0.251, 0.393} & \defscoreci{0.314}{0.248, 0.393} & \defscoreci{0.231}{0.169, 0.305} \\
\shortstack[l]{LoCoMo\\TaskScore $\uparrow$} & \defscoreci{0.256}{0.138, 0.384} & \defscoreci{0.253}{0.138, 0.382} & \defscoreci{0.019}{0.004, 0.035} \\
\bottomrule
\end{tabular*}
\end{table}

The ablation identifies counterfactual comparison, rather than cross-request tracking, as the active component on this subset.
Removing tracking barely changes semantic recovery, HBPS, or TaskScore, whereas removing counterfactual comparison raises UMR-F1 and lowers TaskScore on every benchmark.
This difference matters because suppressing user-related content without asking whether personalization changed the decision can discard useful outputs without isolating the behavior that carries exposure.
The counterfactual test is more selective because it focuses intervention on choices caused by personalization.
This also explains why screening each response for user-related content can miss information exposed through the effect of personalization on a decision.
Even this targeted control releases some personalization-dependent decisions at the evaluated operating point, leaving behavioral evidence from which UMPeek can recover information.

\paragraph{What the defense results imply.}
The three defenses intervene at different points in the release path, and their ordering clarifies which point matters for this attack.
PrivacyChecker evaluates whether a response is appropriate to disclose, and the Theory-of-Mind defense evaluates how a requester might use that response.
Both reason about the released answer.
Stateful Counterfactual instead asks whether personalization changed the decision.
Its lower UMR-F1 indicates that this comparison is better aligned with the evidence UMPeek uses, namely a difference between behavior with and without the retained user information.
The accompanying TaskScore loss also shows why the channel is difficult to close.
The same personalization-dependent difference that supplies attack evidence often supplies the useful adaptation that the application is intended to provide.

This connection changes what a defense must accomplish.
A response that appears harmless alone can become informative after another task rules out a competing explanation, so success on isolated responses does not establish protection over a sequence.
Conversely, suppressing every personalization-dependent choice would reduce exposure by removing personalization itself.
The relevant objective is therefore not only to sanitize text, but to limit how confidently a sequence of released choices can be attributed to retained user information while preserving the choices needed for the task.
The separate UMR-F1, HBPS, and TaskScore results expose this three-way requirement.

\begin{takeaway}
\textbf{Takeaway.}
Response-level defenses screen individual releases, while UMPeek accumulates how personalization changes choices across requests.
Stateful Counterfactual control reduces this channel but leaves residual recovery while lowering personalized task performance.
\end{takeaway}
\section{Related Work}
\label{sec:related-work}

This work connects persistent personalization, direct extraction, behavioral inference, and agent privacy controls.

\paragraph{Personalization and Persistent User State.}

Personalized systems derive profiles, embeddings, compressed contexts, and structured memories for later decisions~\cite{personamemv2,doddapaneni-etal-2024-user,liu-etal-2025-llms,userllm,su-etal-2025-personalized,qiu-etal-2025-latent,ramos-etal-2024-transparent,zhang-etal-2025-prime,zhang-etal-2026-personaagent,mem0}.
Audits of ChatGPT memory find personal and inferred attributes, while persona skills turn histories into portable artifacts with disclosure and impersonation risks~\cite{dash-etal-2026-algorithmic-self-portrait,xiang-etal-2026-persona-skills}.
Key-fact summarization can reduce direct canary extraction~\cite{chen-etal-2026-deployment-time}.
UMPeek instead asks what retained meaning remains inferable without access to history or derived state.

\paragraph{Direct Extraction and Adversarial Manipulation.}

Black-box attacks recover training examples, agent memories, retrieved content or membership, and hidden prompts~\cite{carlini-etal-2021-extracting,mextra,adam,qi-etal-2025-spill-beans,nguyen-etal-2026-five-queries,pleak,llmpbe}.
Recent attacks elicit contextual memory or exfiltrate isolated records through malicious tools~\cite{cui-etal-2026-spore,gao-etal-2026-isolated}.
Exposed embeddings and internal states are also vulnerable to inversion~\cite{morris-etal-2023-text,dong-etal-2025-depth}.
Other attacks manipulate tool actions through adversarial inputs~\cite{injecagent,imprompter}.
UMPeek accesses neither backend state nor tool control, but infers semantic user information from ordinary visible behavior.

\paragraph{Inferring Users from Personalized Behavior.}

Model and attribute inversion infer users from predictions, activity, recommendations, supplied text, or public traces~\cite{fredrikson-etal-2015-model-inversion,gong-liu-2016-attribute-inference,kandpal-etal-2024-user-inference,calandrino-etal-2011-collaborative-filtering,xin-etal-2023-user-behavior-leakage,wang-etal-2025-recsys-inversion,staab2024beyond,pie,du-etal-2026-automated-profile}.
Recent work audits assistant profiling, recovers private negotiation constraints from interaction dynamics, or infers user traits from agent network metadata~\cite{vekaria-etal-2025-big-help,rani-2026-behavioral,jeong-etal-2026-network-agent-leakage}.
UMPeek instead adaptively tests open-ended explanations of inaccessible personalized state through task-dependent choices.

\paragraph{Agent Privacy Evaluation and Defenses.}

Agent audits examine inappropriate use, acquisition, or disclosure of private information in task context~\cite{mireshghallah-etal-2024-confaide,shao-etal-2024-privacylens,zharmagambetov-etal-2025-agentdam,mireshghallah-etal-2026-cimemories,fu-etal-2026-ci,lin-etal-2026-third-eye,zhang-etal-2026-privacypeek,gomaa-etal-2026-converse}.
CMPL uses adaptive conversation to expose failures missed by single-turn audits, while compositional privacy studies how leakage accumulates across responses~\cite{das-etal-2025-beyond,compositional-privacy}.
PrivacyChecker filters contextual flows, SP-Mem separates sensitive values from sanitized memory, and RootGuard protects dependent releases by sanitizing source values once~\cite{privacychecker,wang-etal-2026-spmem,anshumaan-etal-2026-dependency-aware}.
These works protect or audit available private data, whereas UMPeek studies information implied by personalization-dependent choices, including information never stated.

\section{Conclusion}

Transforming a history into a derived user model can remove source wording, while records and backend representations remain inaccessible through the ordinary interface.
UMPeek nevertheless shows that user information retained for personalization can be inferred from the choices it affects.
Hypothesis-guided follow-ups resolve explanations that one response leaves ambiguous.
We observe this mechanism across four benchmarks, three local backends, independently implemented services, an end-to-end personal agent, and the evaluated inference-time defenses.

These evaluations identify a behavioral privacy surface distinct from direct access to records or representations.
UMPeek tests how retained information affects decisions rather than decoding the state that stores it, so this channel does not depend on a particular storage format.
The defense results expose a utility tension.
Reducing the behavioral effect of retained information can lower recovery, but can also remove useful personalization.
Protecting personalized agents therefore requires interaction-level controls that limit what repeated personalization-dependent choices reveal while preserving the adaptation they provide.

\clearpage

\bibliographystyle{IEEEtran}
\bibliography{references}

\appendices

\section{Ethical Considerations}
\label{app:ethical-considerations}

UMPeek has a dual-use risk because a method for detecting behavioral privacy leakage could also be used to infer information about real users.
The principal harm is loss of privacy from choices that do not state the inferred information directly, while testing live services may also consume resources or alter personalized state.
We judged the controlled study warranted because existing evaluations can miss this channel and measuring it can help providers develop interaction-level defenses.

We recruited no participants and accessed no real-user accounts.
The real-world validation used synthetic PersonaMem-v2 histories, while the offline evaluation used released benchmark data with mixed provenance.
PersonaLens derives demographic fields from participants in the PRISM Alignment Dataset and generates the associated preferences and interaction histories~\cite{personalens,prism_alignment}.
The institutional ethics review board of The Hong Kong Polytechnic University approved this use of public secondary data.
We used only released fields, attempted no re-identification, and report aggregate results.

The managed-service experiments used dedicated, author-controlled resources, isolated synthetic identities, documented interfaces, and small request budgets.
We requested permission from the providers but received no response.
Google permits customers to assess their own Cloud projects without prior notification, and AWS lists AgentCore among services that customers may security-test without prior approval, subject to their policies~\cite{google_cloud_security_faq,aws_pentesting_policy}.
The Mem0 evaluation used documented API operations only against state created in our account and did not test provider infrastructure.
We did not access third-party accounts, bypass access controls, inspect hidden provider state, or attempt to recover provider source records, and retained only interface-visible outputs and derived evaluation records.
We will report the findings through the providers' designated security channels.

The released implementation excludes account identifiers, credentials, private platform state, raw external-service records, and third-party source data.
These safeguards reduce immediate harm without eliminating the method's dual-use potential, which motivates the bounded threat model and the defense evaluation reported with the attack.

\paragraph{Artifact availability.}
The code and artifacts accompanying this preprint are available at \url{https://github.com/haoyangliASTAPLE/umpeek/}.
It provides the UMPeek and defense implementations, common interfaces, prompts, configurations, preprocessing and scoring code, compact derived evaluation records, and scripts for rebuilding the released results.
Third-party datasets, credentials, provider resources, and external services' internal state are excluded, with acquisition instructions supplied for public data.
Versioned releases will also be preserved in a permanent archive.

\section{Experimental Details}
\label{app:experimental-details}

\paragraph{Benchmark construction.}
PersonaMem-v2~\cite{personamemv2} contributes 5,000 multiple-choice interactions from 200 personas whose earlier conversations encode implicit preferences.
PersonaLens~\cite{personalens} contributes a fixed subset of 5,000 task-oriented interactions from 79 users.
ETAPP~\cite{etapp} pairs 150 instructions with 32 profiles to yield 4,800 cases in which personalization affects visible tool choices or arguments.
LoCoMo~\cite{locomo} contributes 1,523 question-answer pairs from 20 speakers in 10 multi-session conversations.
The resulting 16,323 targets are evaluated with each of three local personalization backends, yielding 48,969 target-backend interactions.
For unseen-task evaluation, we select three different requests about the same user or conversation and exclude duplicates or cases without scoreable expected behavior.
PersonaLens tasks remain in the same domain when possible, while LoCoMo tasks use related evidence and nearby turns in the same conversation when possible.

\paragraph{Semantic recovery.}
For target $i$, let $R_i$ denote the reference claims, let $\hat{S}_i$ denote the recovered estimate, and let $M_i$ denote the one-to-one matches produced by the rule in Section~\ref{sec:experimental-setup}.
UMR precision is $|M_i|/|\hat{S}_i|$, UMR recall is $|M_i|/|R_i|$, and their harmonic mean is
\[
  \operatorname{UMR\text{-}F1}_i
  =
  \frac{2|M_i|}{|R_i|+|\hat{S}_i|}.
\]
An empty estimate receives zero precision, and reported recovery scores are means of target-level values unless stated otherwise.
UMR-F1 uses the 48,941 target-backend evaluations with at least one reference claim.

\paragraph{Behavioral prediction.}
For every unseen task $h$ associated with target $i$, a fixed rule uses $\hat{S}_i$ to predict behavior $\hat{o}_{ih}$ for request $q_{ih}$ and compares it with reference behavior $o_{ih}$.
The rule selects a supported option for choice tasks, constructs a compatible tool call for ETAPP, and extracts an answer supported by the recovered claims for LoCoMo.
The score $s(\hat{o}_{ih},o_{ih})$ uses exact match for choices, a linearly decreasing rank score for ordered outputs, equal weight on tool-name accuracy and argument F1 for tool actions, and token F1 for free-form answers.
For $N$ targets, the Held-out Behavior Prediction Score is
\[
  \operatorname{HBPS}
  =
  \frac{1}{N}\sum_{i=1}^{N}
  \frac{1}{|H_i|}\sum_{h\in H_i}
  s\!\left(\hat{o}_{ih},o_{ih}\right),
\]
where $H_i$ is the set of unseen tasks paired with target $i$.
Each unseen task is selected so that user-specific information relevant to the target interaction can also affect its reference behavior.

\section{Real-World Validation Details}
\label{app:real-world-details}

\paragraph{Retention and managed-service protocol.}
To evaluate exposure of information that a service actually retained, we first test whether its personalized state contains information semantically equivalent to the target reference claim.
Only targets that pass this check enter the behavior-based comparison, and no attack method receives the retention result.
Google Memory Bank, AWS AgentCore Memory, and Mem0 Platform each contribute 20 retained targets evaluated with all eight methods under a cap of three total requests per target.
Google and Mem0 produce visible behavior with Gemini 2.5 Flash-Lite~\cite{gemini_2_5_flash_lite}, while AWS uses Amazon Nova 2 Lite~\cite{amazon_nova_2_lite}.
All managed-service results were collected in August 2026.

\paragraph{OpenClaw protocol.}
We instantiate 50 OpenClaw agents from synthetic personas and test 206 reference claims for retention.
OpenClaw retains 28 claims across 25 personas, from which we select one retained claim per persona to obtain 25 targets.
These targets are not used to rank the eight methods in the managed-service comparison.

\section{Adaptive Defense Details}
\label{app:defense-details}

\paragraph{Protocol.}
The main comparison samples 128 targets from every benchmark-backend pair and allows the initial interaction plus at most 16 follow-ups.
The retained personalized state, attack procedure, and request cap remain fixed across undefended and defended conditions.
UMPeek selects later requests using only behavior released by the defense.
TaskScore applies each benchmark's original task metric~\cite{personamemv2,personalens,etapp,locomo} to the evaluated personalized tasks.
The threshold study uses 32 targets from every benchmark-backend pair and reuses the same targets at thresholds 0.5, 1.0, and 1.5.
Mechanism ablations use the same subset at threshold 1.0.
For uncertainty estimates, outcomes for the same benchmark target across the three local backends form one resampling cluster, and macro scores are unweighted means over the four benchmarks.

\end{document}